\documentclass[amsmath,amssymb,superscriptaddress]{revtex4}

\usepackage{graphicx}
\usepackage{dcolumn}
\usepackage{bm}
\usepackage{graphicx}
\usepackage{dcolumn}
\usepackage{amsmath}
\usepackage[latin1]{inputenc}
\usepackage{graphicx, psfrag}
\usepackage{amssymb}
\usepackage[colorlinks=true, citecolor=blue, urlcolor = blue, linkcolor= red, bookmarks=true]{hyperref}
\usepackage{float}
\usepackage{amsmath}
\usepackage{amsfonts}
\usepackage{dcolumn}
\usepackage{hyperref}
\usepackage{subfigure}
\usepackage{pgfplots}
\usepackage{epstopdf}
\usepackage{booktabs}

\begin{document}
\title{Noncommutative inspired black holes in regularized 4D Einstein-Gauss-Bonnet theory}
\author{Sushant G. Ghosh}
\email{sgghosh@gmail.com,sghosh2@jmi.ac.in}
\affiliation{Centre for Theoretical Physics, Jamia Millia Islamia,  New Delhi 110025, India}
\affiliation{Astrophysics and Cosmology Research Unit, School of Mathematics, Statistics and Computer Science, University of KwaZulu-Natal, Private Bag 54001, Durban 4000, South Africa}
\author{Sunil D. Maharaj}\email{maharaj@ukzn.ac.za}
\affiliation{Astrophysics and Cosmology
	Research Unit, School of Mathematics, Statistics and Computer Science, University of
	KwaZulu-Natal, Private Bag 54001, Durban 4000, South Africa}
\date{\today}

\begin{abstract}
Low energy limits of string theory indicated that the standard gravity action should be modified to include higher-order curvature terms, in the form of dimensionally continued Gauss-Bonnet densities.
If one includes only quadratic curvature terms then the resulting theory is Einstein-Gauss-Bonnet (EGB) gravity valid only in $D>4$ dimensions. Recently there has been a surge of interest in regularizing, a $ D \to 4 $ limit, of EGB gravity, and the resulting regularized $4D$ EGB gravities have nontrivial gravitational dynamics. We obtain a static spherically symmetric noncommutative (NC) geometry inspired black hole solution with  Gaussian mass distribution as a source in the regularized $4D$ EGB and also analyze their properties. The metric of the NC inspired $4D$ EGB black hole smoothly interpolates between a de Sitter core around the origin and $4D$ EGB metric as $r/\sqrt{\theta} \to \infty$.  Owing to the NC and GB term corrected black hole,  the thermodynamic quantities have also been altered.  The phase transitions for the local thermodynamic stability, in the theory, is outlined by a discontinuity of specific heat  $C_{+}$  at a critical radius $r_+=r_C$, and  $C_{+}$ changes from infinitely negative to infinitely positive and then down to a finite positive for the smaller $r_+$. The thermal evaporation process leads to a  thermodynamic stable extremal black hole with vanishing temperature.  
\end{abstract}
\maketitle

\section{INTRODUCTION}
The Gauss-Bonnet (GB) correction to the Einstein-Hilbert action  contains only the quadratic terms in the curvature tensor and also  appears naturally in the effective action of heterotic string theory \cite{30, 31, 32}, defined by 
\begin{equation}
\mathcal{L}_{GB}=R_{\mu\nu\gamma\delta}R^{\mu \nu\gamma\delta}-4R_{\mu\nu}R^{\mu\nu}+R^{2}.
\end{equation}
EGB  gravity, which also appears in six-dimensional Calabi-Yau compactifications of \textit{M}-theory \cite{33}, is  a natural generalization of  general relativity (GR) to $D\geq 5$, proposed  by Lanczos \cite{Lanczos}, and  David Lovelock \cite{dll}.   The theory allows us to explore how the GB quadratic curvature corrections substantially change the qualitative physical features we know from our understanding of  black holes 
in GR. Boulware and Deser \cite{bd} obtained the first spherically symmetric static black hole solution in EGB gravity, 
(see also \cite{Wiltshire:1988uq} for  charged black holes), and later analysed in cascade of works \cite{Ghosh:2014pga,Sahabandu:2005ma,ms} including the colored black holes solution of Yang-Mills-Dilaton-Gravity system in the presence of a Gauss-Bonnet term \cite{Kanti:1996gs}.  It was also demonstrated that  in the Einstein-scalar-Gauss-Bonnet theory with a coupling function the  no-hair theorems are easily evaded, and also lead to a large number of regular black-hole solutions \cite{Antoniou:2017acq,Bakopoulos:2018nui,Antoniou:2017hxj}

Interestingly, EGB gravity yields conserved second order equations of motion in arbitrary 
number of dimensions $D\geq 5$  which help  us to investigate  several conceptual issues of gravity in a 
more general framework, and the EGB theory is free of ghosts about other exact backgrounds \cite{bd}. 
The GB term, in $D<5$, is a topologically invariant term, i.e.,  its variation is a total derivative and 
thus does not affect the classical  equations of motion. Thus, in $4D$, EGB theory coincides with GR \cite{Lanczos}, 
but in higher dimensions both theories are actually different.

Recently, Glavan and Lin \cite{Glavan:2019inb}  proposed a  regularised $4D$ EGB gravity; they first rescaled the GB coupling,  as $\alpha/(D-4)$, and then took the limit $D = 4$ in the equations of motion. This process results in  
nontrivial contributions from the GB term in the equations of motion in $4D$  spacetime. This regularization procedure was originally proposed by Tomozawa \cite{Tomozawa:2011gp}  with  finite one-loop  quantum corrections to Einstein gravity,  and they also found the spherically symmetric black hole solution  which results in the repulsive nature of gravity at short distances. Later  Cognola {\it et al.} \cite{Cognola:2013fva}   simplified the approach  of Tomozawa \cite{Tomozawa:2011gp} by reformulating the arguments that  mimic quantum corrections due to a GB invariant within a classical Lagrangian approach.

The gravity action \cite{Glavan:2019inb,Cognola:2013fva}, with rescaled coupling constant $\alpha/(D-4)$,  can be written as
\begin{equation}\label{Action}
\mathcal{I}_{G}=\frac{1}{2}\int_{\mathcal{M}}dx^{D}\sqrt{-g}\left[  \mathcal{L}_{1} +\frac{\alpha}{(D-4)} \mathcal{L}_{GB}
\right] + \mathcal{I}_{S},
\end{equation}
with $\kappa =1$. $\mathcal{I}_{S}$ denotes the action associated with matter and $\alpha$ is a
coupling constant with dimension of $(\mbox{length})^2$  which is positive in  heterotic string theory. 
Here, $R_{\mu\nu}$, $R_{\mu\nu\gamma\delta}$ and $R$ are the Ricci tensor, Riemann tensor, and  
Ricci scalar, respectively. The variation of the action with respect to the metric $g_{\mu\nu}$ gives 
the EGB equations \cite{Ghosh:2020syx,Ghosh:2020vpc}
\begin{equation}\label{ee}
R_{\mu\nu} - \frac{1}{2} R g_{\mu\nu}+\frac{\alpha}{(D-4)}  G_{\mu\nu}^{GB}=T_{\mu\nu}^{S},
\end{equation}
where  $G_{\mu\nu}^{GB}$ is explicitly given
by \cite{Ghosh:2020syx,Ghosh:2020vpc}
\begin{eqnarray}
G_{\mu\nu}^{GB} & = & 2\;\Big[ -R_{\mu\sigma\kappa\tau}R_{\quad\nu}^{\kappa
	\tau\sigma}-2R_{\mu\rho\nu\sigma}R^{\rho\sigma}-2R_{\mu\sigma}R_{\ \nu
}^{\sigma} \nonumber \\ & &  +RR_{\mu\nu}\Big] -\frac{1}{2}\mathcal{L} _{GB}g_{\mu\nu}. 
\end{eqnarray}
and $T_{\mu\nu}^{S}$ is the energy-momentum tensor for the matter fields. We note that the divergence of the EGB tensor $G_{\mu \nu}^{GB}$ vanishes. We assume that the metric has the  form \cite{Ghosh:2014pga,ms}
\begin{equation}\label{metric}
ds^2 = -f(r) dt^2+ \frac{1}{f(r)} dr^2 + r^2 \tilde{\gamma}_{ij}\; dx^i\; dx^j,
\end{equation}
where $ \tilde{\gamma}_{ij} $ is the metric of a $D-2$ dimensional constant curvature space with $k = -1,\; 0,\;$ or $1$. 
In this paper, we shall restrict our attention to $k = 1$. 
Let us consider the metric (\ref{metric}) and solving, in the limit $D\to 4$, the $(r,r)$-component of equation (\ref{ee}),  
 which leads to  the static spherically symmetric black hole solution in $4D$ EGB gravity  \cite{Glavan:2019inb,Tomozawa:2011gp,Cognola:2013fva}
\begin{equation}
f(r)= 1 + \frac{r^2}{4 \alpha }
\Biggl[ 1\pm \Big( 1+ \frac{16 \alpha  M }{r^3} \Big)^{1/2} \Biggr], \label{lapse}
\end{equation}
where  $M$ is the integration constant related to the black hole mass. On the other hand, 
in the limit $r \to 0$, the metric function (\ref{lapse}) shows the repulsive nature of quantum 
corrections to gravity at short distances and thus the gravitational potential does not diverge 
at $r=0$ \cite{Glavan:2019inb,Tomozawa:2011gp}  and one can conclude that black holes in this 
regularised $4D$ EGB gravity are singularity-free. Since the gravitational force is 
repulsive at small distances an infalling particle fails to reach  
the singularity \cite{Glavan:2019inb}.  Interestingly the black hole solution (\ref{metric}),
with the metric function (\ref{lapse}), was shown to be a solution of the semi-classical 
Einstein equations with conformal anomaly \cite{Cai:2009ua}.  Nevertheless, recently
interesting measures have been taken to analyse this $4D$ EGB theory, in particular, 
the spherically symmetric black hole solution (\ref{lapse}) was extended to include 
charge for an anti-de Sitter spacetime \cite{Fernandes:2020rpa}, a Vaidya-like 
radiating black hole in Ref.~\cite{Ghosh:2020vpc,Ghosh:2020syx}, black holes 
coupled with magnetic charge \cite{Singh:2020xju,Kumar:2020uyz,Kumar:2020bqf,Kumar:2020xvu}, 
cloud of string models \cite{Singh:2020nwo}, its stability and quasi-normal modes 
\cite{Konoplya:2020bxa,Churilova:2020aca,Mishra:2020gce,Yang:2020czk,Zhang:2020sjh,Aragon:2020qdc}, 
and  also rotating black holes \cite{Wei:2020ght,Kumar:2020owy}. 
Other probes include relativistic star \cite{Doneva:2020ped},  derivation of regularised field equations \cite{Fernandes:2020nbq},  Morris-Thorne like wormholes \cite{Jusufi:2020yus}, accretion disk around black holes \cite{Liu:2020vkh}, thermodynamics \cite{HosseiniMansoori:2020yfj,EslamPanah:2020hoj},  gravitational lensing  by a black hole \cite{Islam:2020xmy,Jin:2020emq,Heydari-Fard:2020sib}, and generalization to more general Lovelock gravity theory \cite{Konoplya:2020qqh}. 

However, several questions \cite{Ai:2020peo,Hennigar:2020lsl,Shu:2020cjw,Gurses:2020ofy,Mahapatra:2020rds} have been 
raised on the above regularization procedure proposed  in \cite{Glavan:2019inb,Cognola:2013fva}, and also some remedies 
have been suggested  \cite{Lu:2020iav,Kobayashi:2020wqy,Hennigar:2020lsl,Casalino:2020kbt}.  
L\"{u} and Pang  \cite{Lu:2020iav} regularized  EGB gravity by compactifying $D$ dimensional EGB gravity 
on $D-4$ dimensional maximally symmetric  space, followed by redefining the coupling as $ \alpha /(D-4)$, 
and then taking the limit $D \to 4$.  The procedure leads to a  well defined special scalar-tensor theory that belongs to the family of Horndeski 
gravity, and is in agreement with the results of \cite{Kobayashi:2020wqy}.  Furthermore  while investigating  $2D$ Einstein gravity, 
several distinct features were pointed out when applying the said approach  \cite{Ai:2020peo}, 
where one may also construct $2D$ black hole solutions \cite{Nojiri:2020tph}.  Hennigar  {\it et al.} \cite{Hennigar:2020lsl}
proposed a well defined $D \to 4$ limit of EGB gravity  generalizing the previous work of  
Mann and Ross \cite{Mann} in obtaining the $D \to 2$ limit of GR. However, the spherically symmetric $4D$  black hole  
solution (\ref{lapse})  remains valid in these regularised theories \cite{Lu:2020iav,Hennigar:2020lsl,Casalino:2020kbt}, 
but not beyond spherical symmetry \cite{Hennigar:2020lsl}.  

The  EGB  theory got pronounced attention when it was shown that the GB term naturally appears as the in the heterotic string effective action. whereas the noncommutative (NC) geometry found genuinely from the study of open string theories. In particular, NC black holes are involved in the study of string and M-theory \cite{Haro}.  The NC spacetime was first investigated by Snyder \cite{snyder} in the context of divergences arising in relativistic quantum field theory. The  main aim of this paper is to search   for $4D$  black holes in the regularized EGB theory with a static, spherically  symmetric Gaussian mass distribution or   NC inspired  analogue of the $4D$ EGB solution  (\ref{lapse}).

The remainder of this paper is organized as follows. We give a review on NC inspired black holes and find a 
NC inspired  static, spherically symmetric black hole solution for the regularised $4D$  EGB gravity in Sec.~\ref{NCEGB}. In Sec.~\ref{EGB-thermodynamics}, we discuss the various  thermodynamic properties of the obtained solution. The investigation of thermodynamical stability and phase transitions are the subject of Sec.~\ref{NCPT}. The paper ends with summarizing ours main findings in Sec.~\ref{concluding}. 

We follow the metric signature ($-,\;+,\;+,\;+,$) and use the natural units in which $8\pi G = c = 1$.

\section{NC inspired Einstein-Gauss-Bonnet black holes}\label{NCEGB}
The NC geometry leads to a smearing of matter distributions, therefore, NC inspired models could lead to a deeper understanding of gravity at high-energy scale. In the NC theory the spatial coordinates can be thought of as operators, which failed to commute, and the resulting commutation relation reads
\begin{equation}
 \left[x_{\mu}, x_{\nu}\right] = i \theta_{\mu\nu}, \label{NC1}
\end{equation}
with $\theta_{\mu\nu} $ is a, real, $D \times D$ 
anti-symmetric matrix which infers the NC manifold and thereby determines the fundamental cell discretization  of spacetime. 
The commutation relation (\ref{NC1}) leads in direction of the resulting uncertainty relation 
\begin{equation}\label{up}
\Delta x^{\mu}\Delta x^{\nu}\geq \frac{1}{2}|\theta^{\mu\nu}|.
\end{equation}
This offers an effective way for the semi-classical gravity and a way to understand the short-distance or high-energy scale behavior of gravity. Hence progress in understanding the effects of NC gravity can be made by
formulating a model in GR. This idea was realized by Nicolini {\it et al.} \cite{Nicolini:2005vd} to proposed one of the first NC geometry inspired  Schwarzschild black holes, which is an exact solution from Einstein equations for the static, spherically symmetric, Gaussian-smeared matter source. The NC geometry inspired  Schwarzschild black holes,  unlike GR, have no curvature singularity at the center, rather a regular de Sitter core. Subsequently, based on this idea, there has been intense activity in the investigation of NC geometry inspired black holes. The model was generalized to the charged case \cite{Ansoldi:2006vg}, also extended to higher-dimensional spacetime \cite{Rizzo:2006zb, Spallucci:2009zz,Gingrich:2010ed}, and then to BTZ black holes \cite{Kim:2007nx} (see also ~\cite{Nicolini:2006}, for a review). However, one can expect that the NC inspired black hole in  EGB gravity may have some different properties 
because of quadratic curvature corrections.

Thus,  due to the validity of the spherically symmetric 
solution (\ref{lapse})  in  various regularised $4D$ EGB gravities  \cite{Lu:2020iav,Hennigar:2020lsl,Casalino:2020kbt}, 
and other \cite{Cai:2009ua,Tomozawa:2011gp} gravities, it is pertinent to consider the NC version of the solution  (\ref{lapse}).  In particular, we explicitly  shall bring out how  the effect of NC  can alter $4D$ EGB black hole solutions  (\ref{lapse})  and their  thermodynamical properties.  In the NC inspired models one can consider the smearing out of conventional mass distributions and its effect on gravity has been  incorporated in the matter. Hence, we can replace the singular mass distribution of a point mass $\mu$ by a Gaussian-smeared matter source, which in $D-$dimensions reads as \cite{Spallucci:2009zz,Rizzo:2006zb}
\begin{equation}\label{rho}
\rho_{\theta}(r) = \frac{\mu }{(4 \pi \theta)^{(D-1)/2}} e^{{-r^2}/{(4\theta)}}.
\end{equation} 
Here, $\theta$ is the NC parameter and has the dimension of length squared, such that $\sqrt{\theta}$ determines the scale for mass distribution diffusion $\mu$, and can be considered to be of the order of Planck length. Thus instead of a singular mass distribution at $r=0$, now we have a smearing over the length scale $\sqrt{\theta}$ given by a Gaussian distribution \cite{Nicolini:2005vd,Spallucci:2009zz,Rizzo:2006zb}. Using the condition $g_{tt}=1/g_{rr}$ and Eq.~(\ref{rho}), the components of the stress energy tensor are given by 
\begin{eqnarray}\label{cemt}
T^0_0 &=& - T^r_r=  \rho_{\theta}(r) = \frac{\mu }{(4 \pi \theta)^{(D-1)/2}} e^{{-r^2}/{(4\theta)}}.
\end{eqnarray}
Next we use the  Bianchi identity  $T^{ab}{}_{;b}=0$ \cite{Rizzo:2006zb}, to  obtain 
the other components of the the energy momentum tensor as 
\begin{eqnarray}\label{EMT}
T^{\theta}_{\theta} &=& T^{\phi}_{\phi} = \rho_{\theta}(r) + \frac{r}{(D-2)} 
\partial_r \rho_{\theta}(r), 
\end{eqnarray}
and the energy momentum tensor  is completely specified by (\ref{cemt}) and (\ref{EMT}) and corresponds 
to an anisotropic fluid.  Now solving the field  Eqs.~(\ref{ee}) for the matter source,  
in the limit $D\to 4$, we obtain a general solution
\begin{equation} \label{sol:egb}
f_{\pm}(r) = 1+\frac{r^{2}}{4{\alpha}}\left[1\pm \sqrt{1+\frac{32\alpha M}{ r^3 \sqrt{\pi}} 
\gamma\left(\frac{3}{2}, \frac{r^2}{4\theta}\right) }\right],
\end{equation}
by appropriately relating $\mu$ with  integrating constants  and 
$\gamma\left(3/2, \, r^2/4\theta\,  \right)$ is the lower incomplete Gamma function \cite{Spallucci:2009zz}
given by
\begin{equation}
\gamma\left(\frac{3}{2}, \frac{r^2}{4\theta}\right)\equiv \int_0^{r^2/4\theta}\; du\; u^{1/2}\; e^{-u}.  
\end{equation}
 Asymptotically far away $\rho_{\theta}=0$.  Equation (\ref{sol:egb}) is an exact solution of the field equation (\ref{ee}) for matter source (\ref{EMT}),   which in the limit  ${r/\sqrt{\theta}} \rightarrow \infty$ reduces to the $4D$ EGB solution (\ref{lapse}).  
On physical grounds, a non-vanishing radial pressure is needed to balance the inward gravitational pull, preventing droplet to collapse into a matter point. This is the basic physical effect on matter caused by spacetime noncommutativity  at distance scale of order $ \sqrt{\theta}.$  Thus, at short distances, significant changes are expected due to NC.

Thus the metric (\ref{metric}) with (\ref{lapse}) describes a NC inspired  black hole in the various 
regularized $4D$ EGB theories \cite{Glavan:2019inb,Hennigar:2020lsl,Lu:2020iav,Kobayashi:2020wqy}.
We define the mass-energy $m(r)$ by
\begin{equation}\label{mass}
m(r) = {2  M } \gamma\left(\frac{3}{2}, \frac{r^2}{4\theta}\right),
\end{equation}
whereas the total mass-energy, $M$, is measured by asymptotic observer \cite{Ansoldi:2006vg}.
We have obtained two branches of solutions, namely, $f_+$ and $f_-$, which correspond to $\pm$ signs in 
front of the square root term. The negative branch of the solution (\ref{sol:egb}), in the limit $\alpha \rightarrow 0$, reduces to
\begin{equation}\label{grsol}
f_{-}(r) = 1-\frac{4 M}{ r \sqrt{\pi}} 
\gamma\left(\frac{3}{2}, \frac{r^2}{4\theta}\right),
\end{equation}
which is exactly the same as the NC inspired $4D$ black hole in GR   \cite{Spallucci:2009zz,Rizzo:2006zb}. 
However, the positive branch, $f_+$, does not converge to GR solution and, henceforth, 
we will consider only the negative branch. Here we note that for $r \ll \sqrt{\theta}$,  $\rho_{\theta} = \rho_{\theta}(0)$ 
whereas for $r \gg 2M$, we have $\rho_{\theta} = 0$.
As $r \to 0 $, the solution of Eq.~(\ref{sol:egb}) becomes
\begin{equation} \label{sol:egb1}
f(r) = 1 -  \sqrt{\frac{2 M r}{ \alpha \sqrt{\pi}} 
	\gamma\left(\frac{3}{2}, \frac{r^2}{4\theta}\right) }.
\end{equation}
This is equivalent to a repulsive core at the short distance.    
\begin{figure*}
	\begin{center}
\begin{tabular}{c c }
\includegraphics[width=0.5\linewidth]{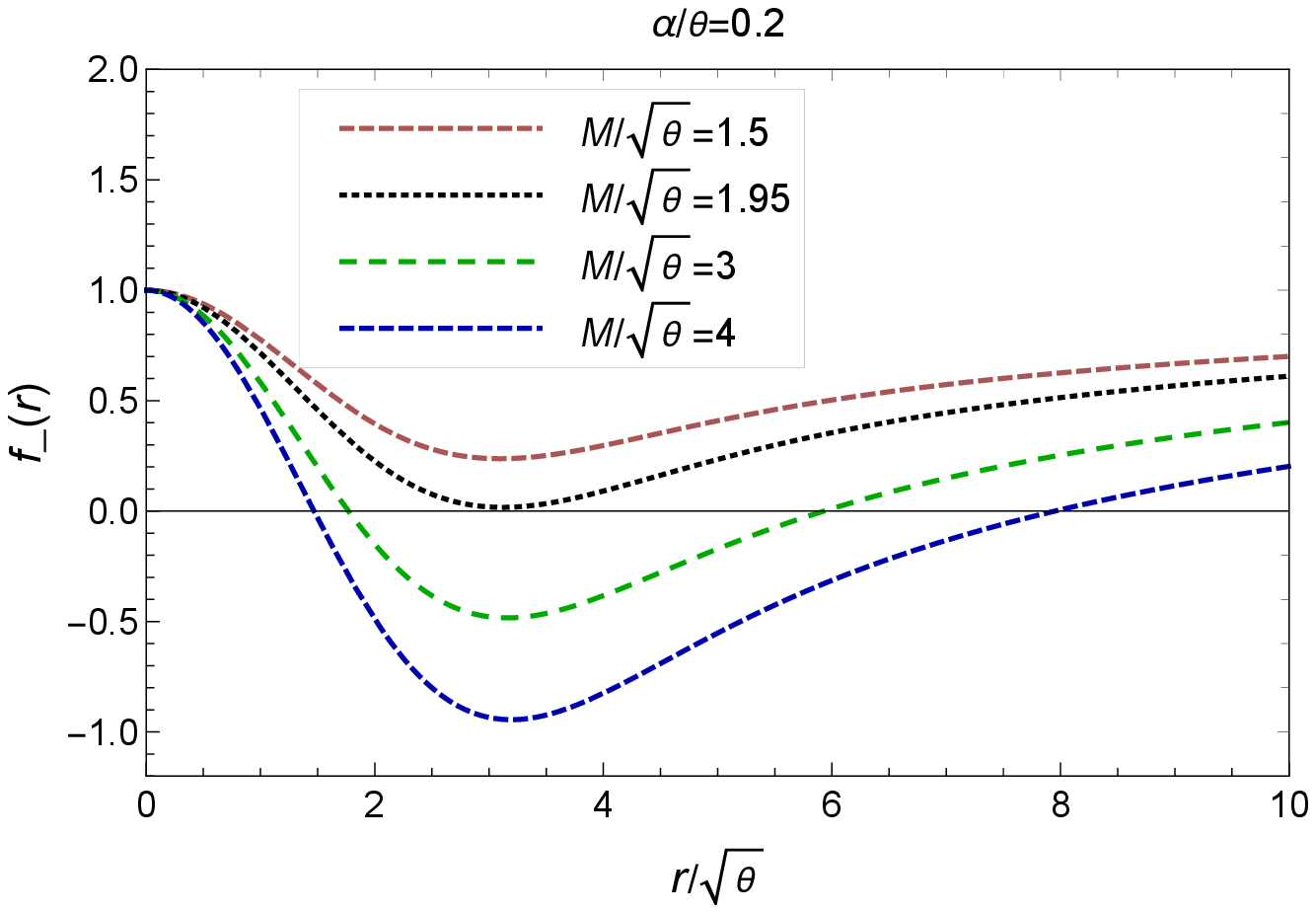}
\includegraphics[width=0.5\linewidth]{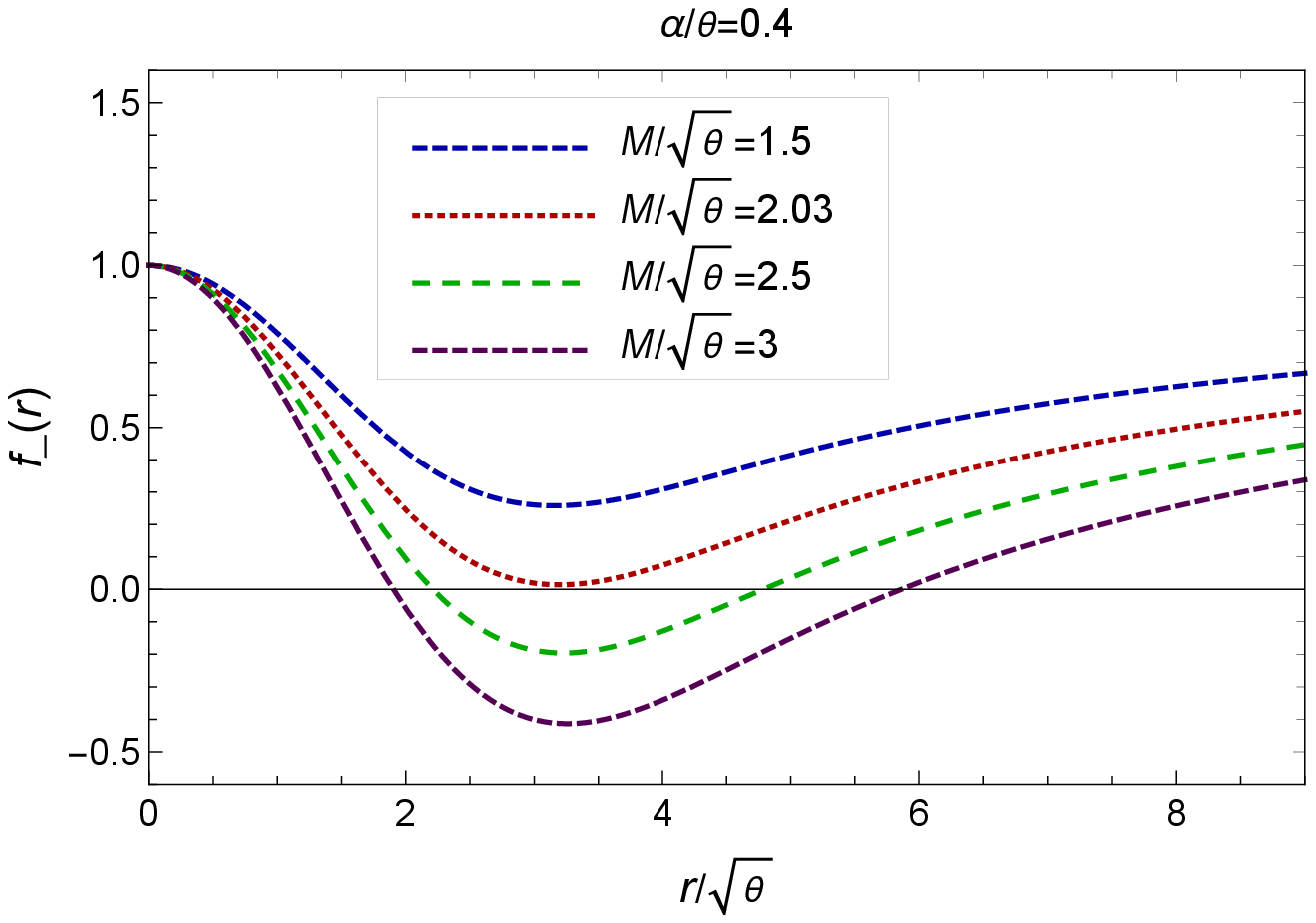}
\end{tabular}
\caption{\label{f4d} Plot of metric function \text{bf $f_(r)$} vs $r/\sqrt{\theta}$, for various values of 
$M/\theta$ for  the NC inspired   4D EGB black holes.}
	\end{center}
\end{figure*}

\subsection{Horizons}
The horizons of this $4D$ EGB black hole are given by  roots of the equation $g^{rr}(r)=0 \,\; \mbox{or} \,\;  f(r)=0$, which implies
\begin{equation}\label{eh}
\sqrt{\pi}	r_{\pm} =  2 M \gamma\left(\frac{3}{2}, \frac{r^2}{4\theta}\right)  \pm \sqrt{  \left[2 M \gamma\left(\frac{3}{2}, \frac{r^2}{4\theta}\right) \right]^2-2\pi \alpha.  }
\end{equation}
These two cases (	$r_{\pm}$ )corresponds, respectively, a
regular non-extremal black hole with a Cauchy horizon ($r_{-}$) and an event horizon ($ r_{+} $).  
 Obviously Eq.~(\ref{eh}) may not be solved exactly 
and hence, we depicted $f(r)$ in Fig.~\ref{f4d}, such that its zeros gives two possible horizons.  It turns out that for a given $\alpha$, there exists a critical 
value of mass  $M$, $M_C$, and critical horizon radius $r_{+}$, $r_C$, given by
\begin{equation}\label{MC}
M_C = \frac{\sqrt{\alpha\pi}}{\sqrt{2}\gamma\left(\frac{3}{2}, \frac{r_C^2}{4\theta}\right) },
\end{equation}
such that $g^{rr}(r_C)=f(r_C)=0$ admits a double root $r_C$, which corresponds to an 
extremal black hole with degenerate horizons.  When $M>M_C$, $f(r)=0$ has two simple 
zeroes,  (cf. Fig.~\ref{f4d}),  corresponding to a nonextremal black hole with  
two horizons viz., a Cauchy horizon (CH) and an event horizon (EH). Whereas $f(r)=0$ has no zeroes for $M<M_C$, i.e., no black hole.   
It is worthwhile to mention that the critical values of $M_C$ and $r_C$ are $\alpha$ dependent, e.g., for
$\alpha=0.2,\;0.4$, respectively $M_C=1.9 \sqrt{\theta},\; 2.03 \sqrt{\theta}$ and
$r_C=3.1156,3.19655$ (cf. Fig.~\ref{f4d}). Figure~\ref{f4d} clearly infers that both critical mass $M_C$ and critical radius $r_C$ increase with increasing GB coupling parameter $\alpha$.  Also, for any value of $\alpha$, one have  $r_+\geq r_C$.  

In the long distance limit, the effect of the NC can be neglected, i.e., in the 
limit  $r/\sqrt{\theta} \rightarrow \infty$, the black hole horizon is located at
\begin{equation}\label{eh1}
	r_{\pm} =   M   \pm \sqrt{  M-2 \alpha }.
\end{equation}
One recovers the horizons of the $4D$ EGB black holes \cite{Glavan:2019inb}, and  for $0<M< 2\; \alpha$, 
we do not have naked singularity \cite{Ghosh:2014pga}.   In the limit 
$\alpha \rightarrow 0$ and $r/\sqrt{\theta} \rightarrow \infty$,  
Eq.~(\ref{eh}) gives $4D$ Schwarzschild black hole  horizon $r_+=2M$.  

\begin{figure*}
	\begin{tabular}{c}
	\includegraphics[width=0.55\linewidth]{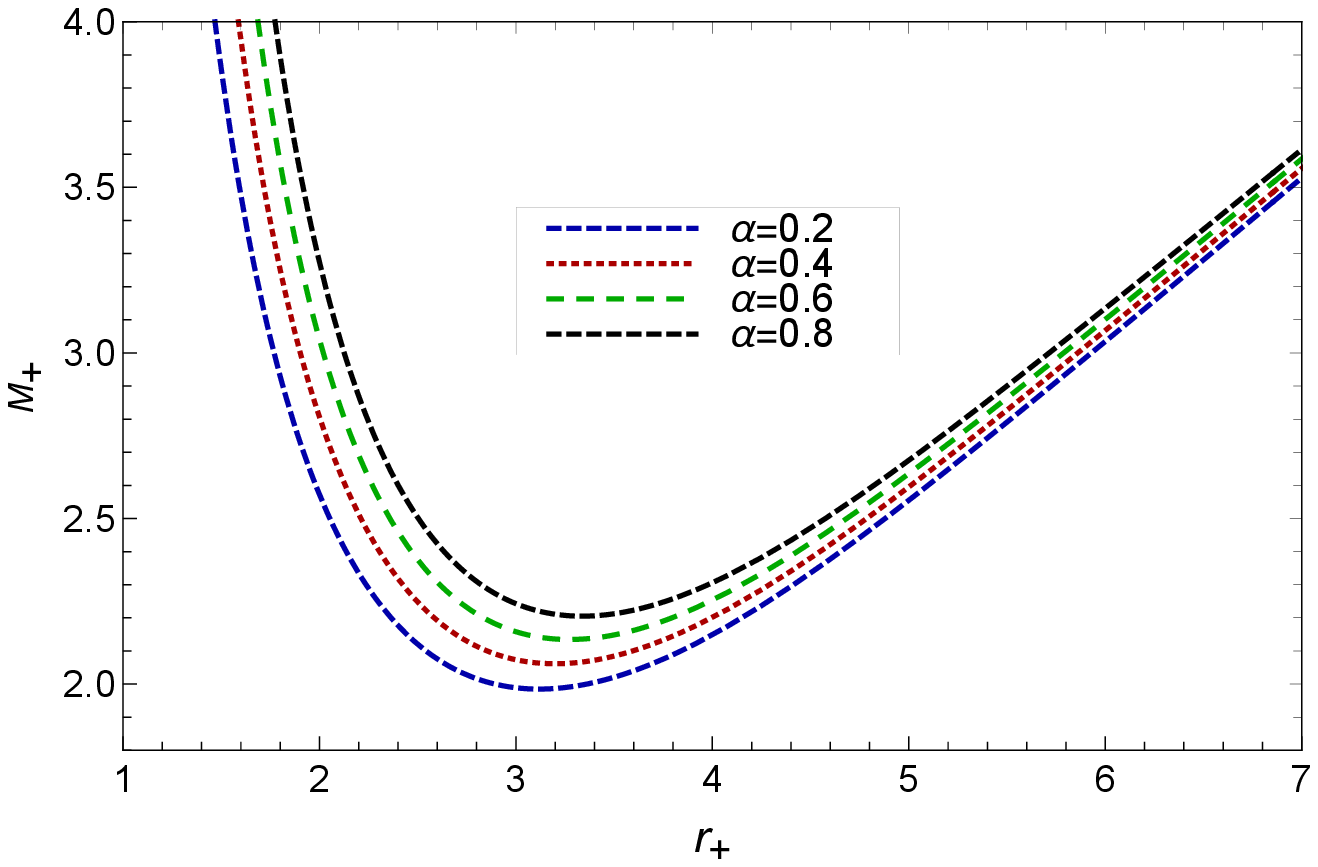}
	\includegraphics[width=0.55\linewidth]{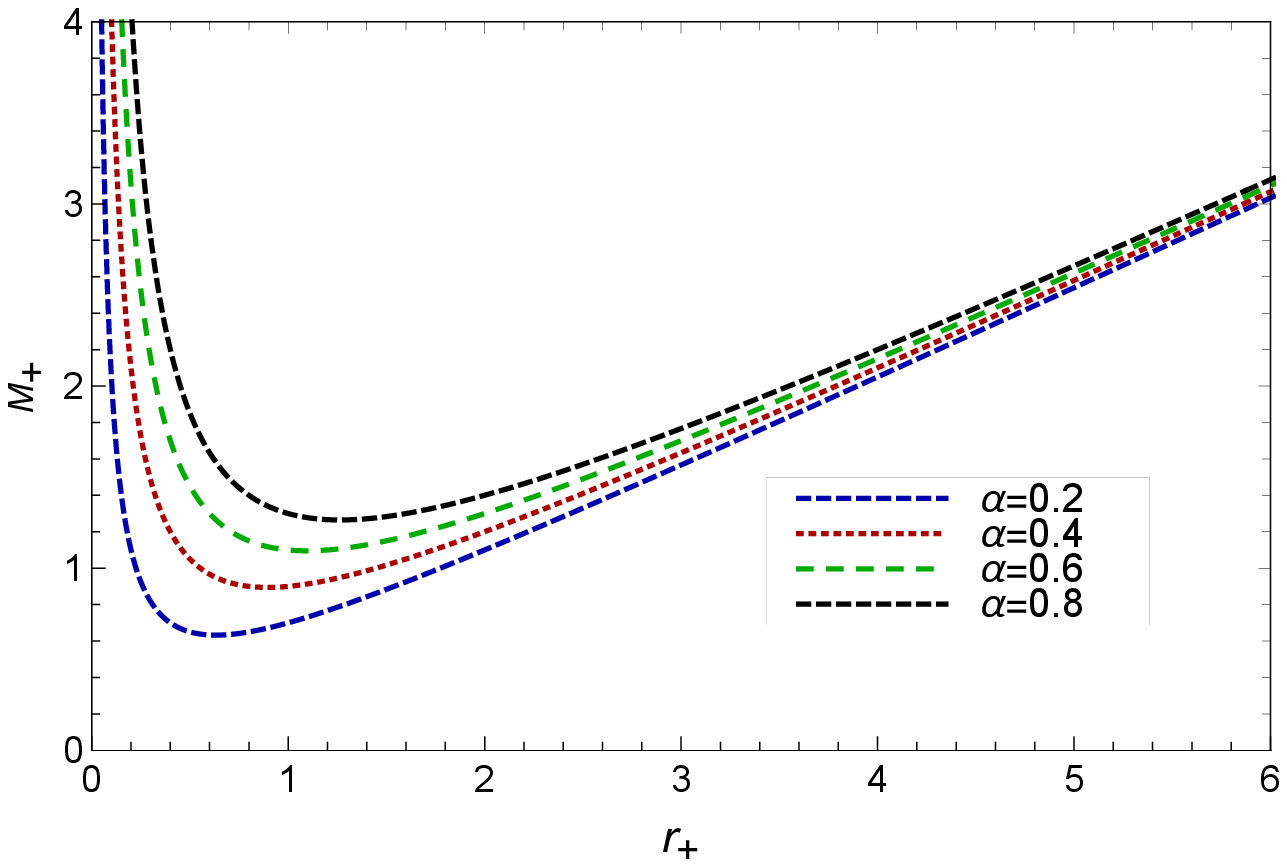}
	\end{tabular}
	\caption{\label{hmass} The black hole mass ($ M_+ $) vs horizon radius $r_+$ for different values of $\alpha$,  
	for the NC inspired   4D EGB black hole (left), which is compared with the commutative counterpart (right).  }
\end{figure*}

\section{Black hole thermodynamics}\label{EGB-thermodynamics}
It may be important to understand how the NC affects the thermodynamical properties of the  NC inspired $4D$ EGB black holes; henceforth we restrict only to the negative 
branch of the solution (\ref{sol:egb}). NC inspired $4D$ EGB black holes are characterized by their mass $M$, GB coupling parameter $\alpha$, and the NC parameter $\theta$. The black hole mass can be expressed in terms of its horizon radius and other parameters, solving (\ref{sol:egb}) leads to
\begin{eqnarray}\label{M1}
	M_{+} =  \frac{\sqrt{\pi}( r_{+}^2+2\alpha)}{4 r_{+} \left[\gamma\left(\frac{3}{2}, \frac{r_{+}^2}{4\theta}\right)\right]}, 
\end{eqnarray}
which in the limit $\alpha \to 0$ and $r/\sqrt{\theta} \rightarrow \infty$ reduces to the $4D$ Schwarzschild black hole mass $M_{+}=r_+/2 $. Let us analyse the  $M_{+}$ to the smearing due to NC effects, which is depicted in the 
left panel of Fig.~ \ref{hmass}, where we  shown  $M_{+}  $  as a function of horizon radius $r_+$ 
for different values of $\alpha$ and compared  with the commutative counterpart on the right panel 
of Fig.~ \ref{hmass}.   As expected,  in the asymptotic large  $r_+$,  both cases  yield essentially 
the same results. Also in both cases a minimum mass $ M_{+}^{min} $ occurs at relatively small $r_{+}$  
but mass  $ M_{+}^{min} $ values are larger in the NC case and so is the radius $r_{+}^{min}$ where minimum mass appears.   
$ M_{+}^{min} $ increases and the corresponding value of $r_{+}^{min}$ increases in both cases as $\alpha$ is increased. 
However, in the region of $r_{+}<r_{+}^{min}$ where NC effects is evident, one can observe notable differences 
in shape of the  two BH mass distribution (cf. Fig.~ \ref{hmass}).

\begin{figure*}
	\begin{tabular}{c c}
		\includegraphics[width=0.55\linewidth]{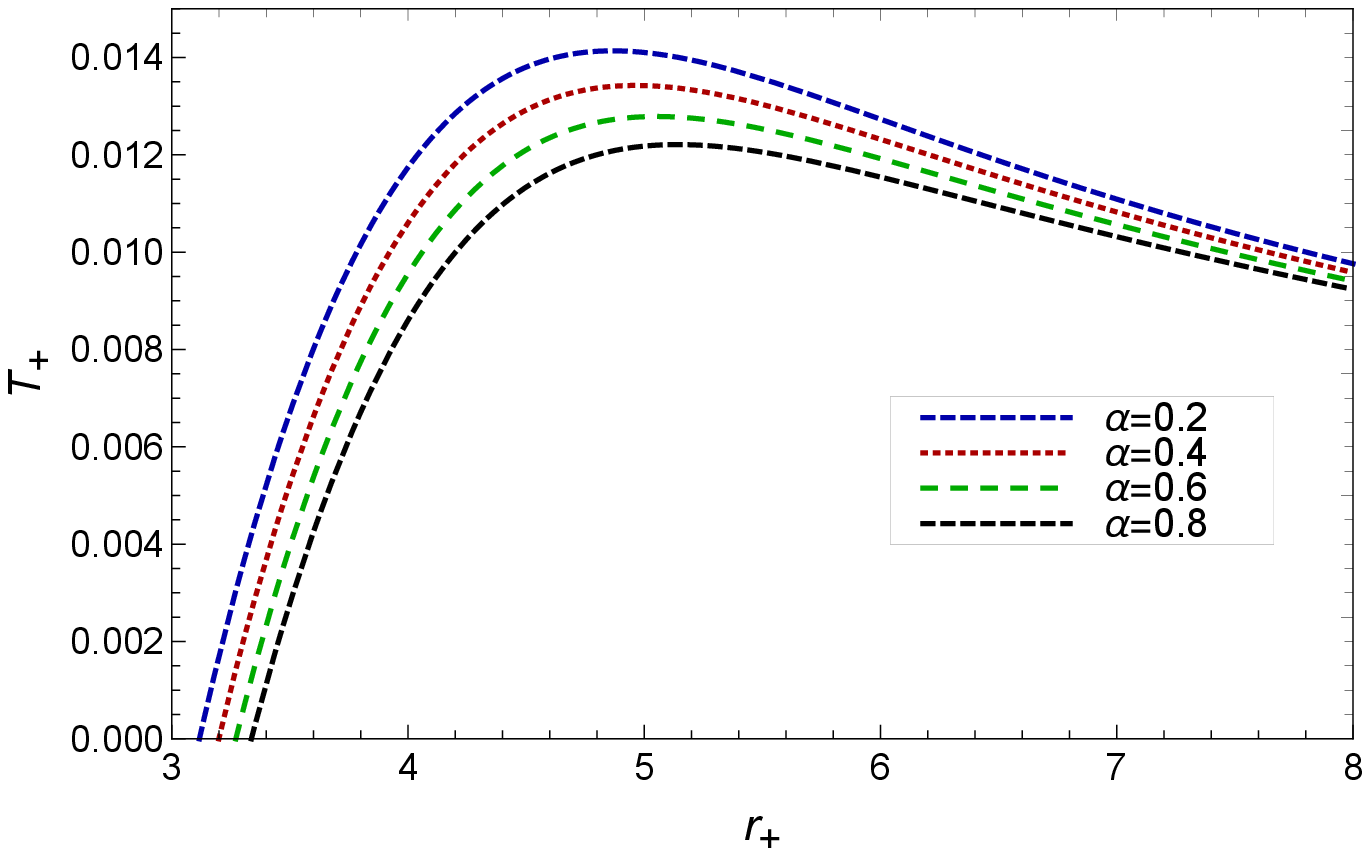}
		\includegraphics[width=0.55\linewidth]{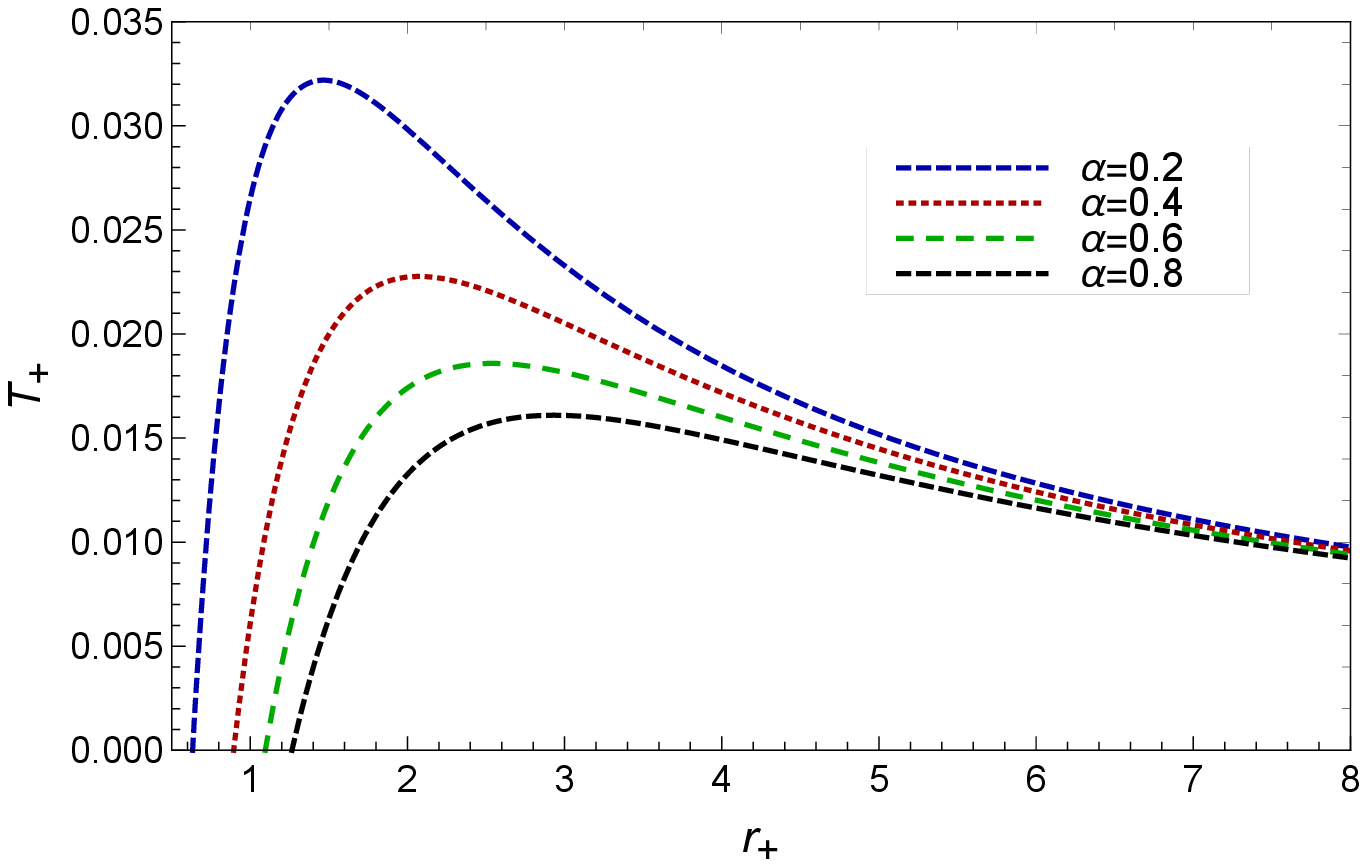}
	\end{tabular}
	\caption{\label{fig:egb:T} The Hawking temperature ($ T_+ $) vs horizon radius $r_+$ 
	for different values of $\alpha$,  for the NC inspired   4D EGB black hole (left)  
	which is compared with the commutative counterpart (right).  }
\end{figure*}

The Hawking temperature can also help us to  understand  the final stage of the  NC 4D EGB black hole evaporation. 
The Hawking temperature associated with the  black hole is defined by $T=\kappa/2\pi$, where $\kappa$ is 
the black hole surface gravity  \cite{Sahabandu:2005ma,Ghosh:2014pga} defined by
\begin{equation}
\kappa^{2}=-\frac{1}{4}g^{tt}g^{ij}g_{tt,i}\;g_{tt,j},
\end{equation}
which on using the black hole metric function, simplifies to
\begin{equation}
\kappa=\left\vert \frac{1}{2}f^{\prime}(r_+)  \right\vert.
\end{equation}
Accordingly, using the metric function (\ref{sol:egb}), the NC inspired $4D$ EGB black hole horizon  temperature reads as 
\begin{eqnarray}\label{temp1}
T_+  &=&  \frac{1}{4 \pi(r_+^2 + 4 \alpha)} \left[\frac{{r_{+}^2-2 \alpha}}{r_{+}}- (r_+^2 + 2 \alpha)
\frac{\gamma' \left(\frac{3}{2}, \frac{r_{+}^2}{4\theta}\right)}{\gamma\left(\frac{3}{2}, \frac{r_{+}^2}{4\theta}\right)}\right].  
\end{eqnarray}
From Eq.~(\ref{temp1}), in the limit 
$r/\sqrt{\theta}\rightarrow0$, we recover temperatures of  the analogous commutative  black holes which are
exactly the same as those obtained in \cite{Singh:2020nwo}.  
We show the NC black hole Hawking temperature ($T^+$) as a function of $r_{+}$ in the left panel of Fig.~\ref{fig:egb:T} for different values of $\alpha$  and compare with the commutative counterpart in right panel. We observe that the $T^+$, in both cases, goes through a maximum as $r_{+}$ decreases and then falls to zero at the BH minimum mass point.  Also, when  $\alpha \neq 0$, the  Hawking temperature ($T^+$) has a peak which decreases and shifts to left  as $\alpha$ increases (cf. Fig.~\ref{fig:egb:T}). The maximal Hawking temperature ($T_+^{Max}$)  occurs at a critical radius $r_C^T$. 
It turns out that the $T_+^{Max}$  decreases with $\alpha$, while $r_C^T$ increases.  
In the region $r_{+} = \sqrt{\theta}$, for $\alpha = 0.2\,  \theta,\; 0.4 \, \theta,\;  0.6 \, \theta, \; 0.8 \,  \theta$, respectively, the temperature ($T^+$), unlike in classical GR  deviates from the increasing nature,  instead of exploding with shrinking $r_{+}$, ($T^+$) reaches a maximum  $T_+^{Max} = 0.141361 \; \sqrt{ \theta },\;  0134251\; \sqrt{ \theta },\;012757 \; \sqrt{ \theta }, \;0.0122104 \; \sqrt{ \theta }$ at $r_{+} = 4.86647\; \sqrt{ \theta },\; 4.9622\; \sqrt{ \theta },\; 5.05769\; \sqrt{ \theta },\; 5.14095 \sqrt{ \theta }$, and then quickly drops to zero at 
$r_{+} = 3.11708 \; \sqrt{ \theta },\; 3.19874 \; \sqrt{ \theta },\; 3.27121 \; \sqrt{ \theta },\;3.33682 \; \sqrt{ \theta }$, where  horizons becomes degenerate, and we are left with an extremal black hole with minimal mass as a stable remnant  and  it has an increasing size when $\alpha$ takes larger values.  We note that  $T_+=1/4\pi r_+$ is the Hawking temperature of the  Schwarzschild  black hole \cite{Ghosh:2014pga}, 
in the limit $\alpha \rightarrow 0$ and $r_+/\sqrt{\theta} \rightarrow \infty$, which can 
be deduced from  Eq.~(\ref{temp1}), and $T_+ \to \infty$ as $r_+ \to 0$.  However,  
this divergence behaviour of the temperature is resolved  in the   $4D$ EGB black holes  \cite{Singh:2020nwo} (Fig.~\ref{fig:egb:T}),  and the same is true for the analogous  NC case as shown in Fig.~\ref{fig:egb:T}. 

Next, we discuss the entropy of NC inspired $4D$ EGB black hole, which can be obtained by the first law of black hole thermodynamics. In GR, entropy satisfies the black hole's area law, which states that the entropy of a black hole is a quarter of the event horizon area \cite{Cai:2003kt,Sahabandu:2005ma,ms}. 
The first law of black hole thermodynamic reads as \cite{Cai:2003kt,Sahabandu:2005ma,Ghosh:2014pga}
\begin{equation}\label{flaw}
dM_+ = T_{+}dS_{+}.
\end{equation}
Hence, the entropy can be obtained from the integration 
\begin{eqnarray}\label{ent:formula}
S_{+} = \int {T_+^{-1} dM} = \int {T_+^{-1}\frac{\partial M_+}{\partial r_+} dr_+}, 
\end{eqnarray}
and using the black hole mass Eq.~(\ref{M1}) and horizon temperature Eq.~(\ref{temp1}) into (\ref{ent:formula}), the entropy of the 
NC inspired EGB gravity black holes is given by 
\begin{eqnarray}\label{S:EGB}
S_{+} = 4 \pi \sqrt{\pi}\int  \frac{4\alpha+r_{+}^2}{r_{+} \gamma\left(\frac{3}{2}, \frac{r_{+}^2}{4\theta}\right)}d r_+.
\end{eqnarray}
The entropy for our model, in the limit 
$r/\sqrt{\theta}\rightarrow 0$,   integrates  to  
\begin{equation}\label{S1}
S_{+} = \frac{A}{4} +  4 \alpha \log \frac{A}{A_0},
\end{equation}
exactly the same as in  Refs. \cite{Singh:2020nwo,Fernandes:2020rpa} with $A=\pi r_{+}^2$ , which has a logarithmic correction. Here, $A_0$ integration constant with units of area. Entropy Eq. (\ref{S1}) is modified due to  parameter   $\alpha$, and the identification between entropy and area is no longer valid for $4D$ EGB black holes.
Notice that, in limit $\alpha \to 0$, we obtain the entropy  of the $4D$ Schwarzschild black hole \cite{Singh:2020nwo}, and the area law holds.

\section{Thermodynamic Stability }\label{NCPT}
In this section, we analyse the thermodynamic stability of the NC inspired $4D$ EGB black holes which requires the  study  of its heat capacity, which is  defined as  in \cite{Cai:2003kt,Ghosh:2014pga}:
\begin{equation}\label{sh_formula}
C_+ = \frac{\partial{M_+}}{\partial{T_+}}= \left(\frac{\partial{M_+}}{\partial{r_+}}\right)
\left(\frac{\partial{T_+}}{\partial{r_+}}\right)^{-1}. 
\end{equation} 
On using Eqs.~ (\ref{M1}) and (\ref{temp1}) we obtain
\begin{widetext}
\begin{eqnarray} \label{egb:sh}
&& C_+ = \frac{( \pi)^{3/2} (r_{+}^2+4\alpha)^2 \left[ \beta r_{+} \gamma' - (\beta- 4 \alpha) \gamma\right]}{r_{+}^2 (\beta+2\alpha) 
\beta\, \gamma\, \gamma{''} - \zeta \, r_{+}^2 \gamma'^2 + 4 \alpha r_{+}^3 \gamma \gamma' + 
(r_{+}^2-10 r_{+}^2 \alpha-8 \alpha^2)\gamma^2}, \nonumber \\ && \mbox{with}\,\,\,\, \gamma = \gamma\left(\frac{3}{2}, \frac{r_{+}^2}{4\theta}\right),\; \beta = (r_{+}^2+2 \alpha) \; \mbox{and} \;  \zeta =  (r_{+}^4 +6\alpha r_{+}^2+8 \alpha^2)
\end{eqnarray}
\end{widetext}
 It turns out that  at some stage, a black hole, due to thermal fluctuations, 
 absorbs more radiation than it emits which leads to positive  heat capacity. 
 Whereas  when the black hole emits more radiation than it absorbs, the heat capacity becomes 
 negative. If the heat capacity  is positive (negative), then the black hole is stable (unstable)  
 to the thermal fluctuations \cite{Cai:2003kt,Sahabandu:2005ma,Ghosh:2014pga}.  
 Here, we analyse this situation for the NC inspired $4D$ EGB black holes. 
 
\begin{figure*}
	\begin{tabular}{c c}
		\includegraphics[width=0.55\linewidth]{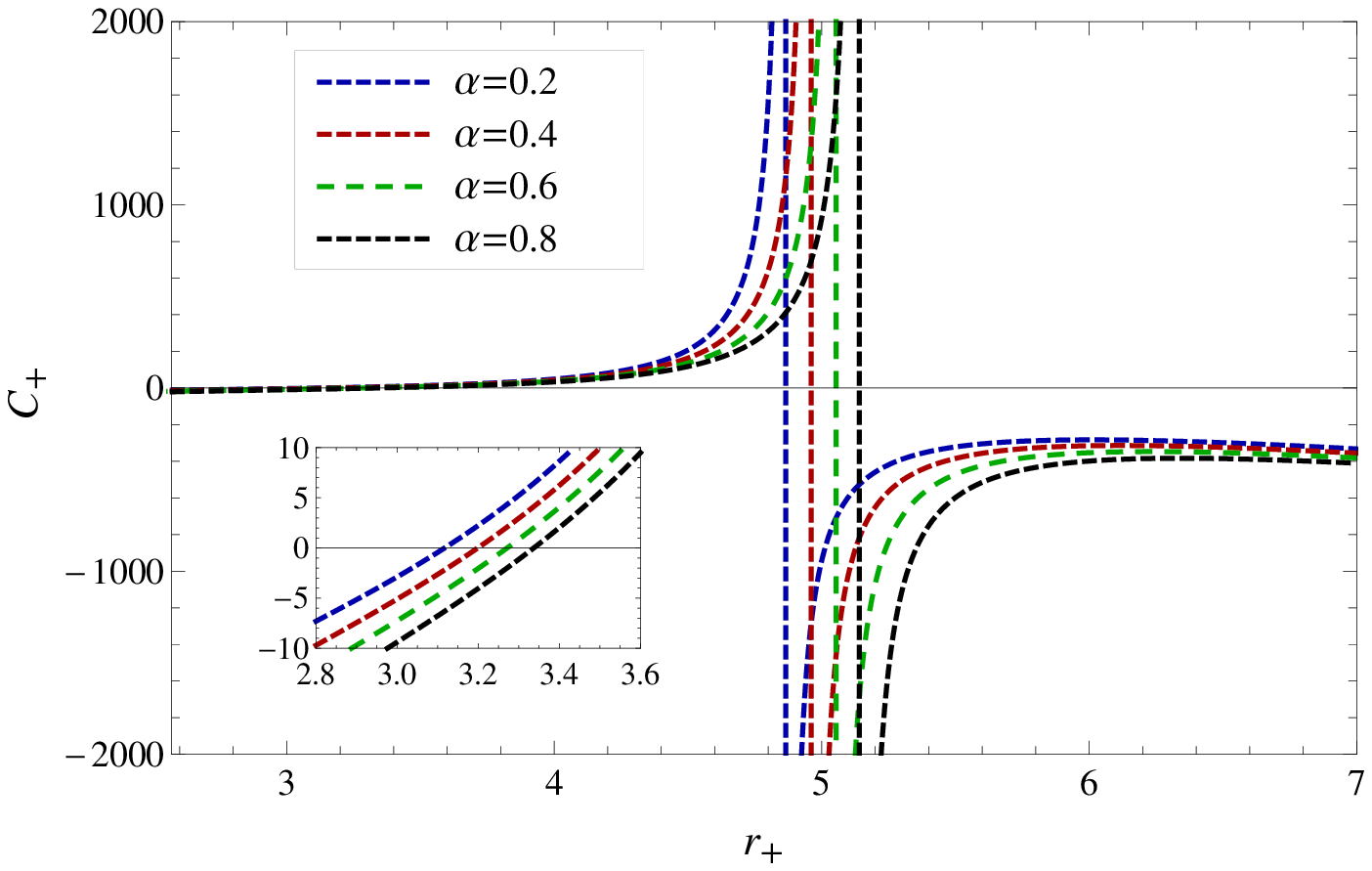}
		\includegraphics[width=0.55\linewidth]{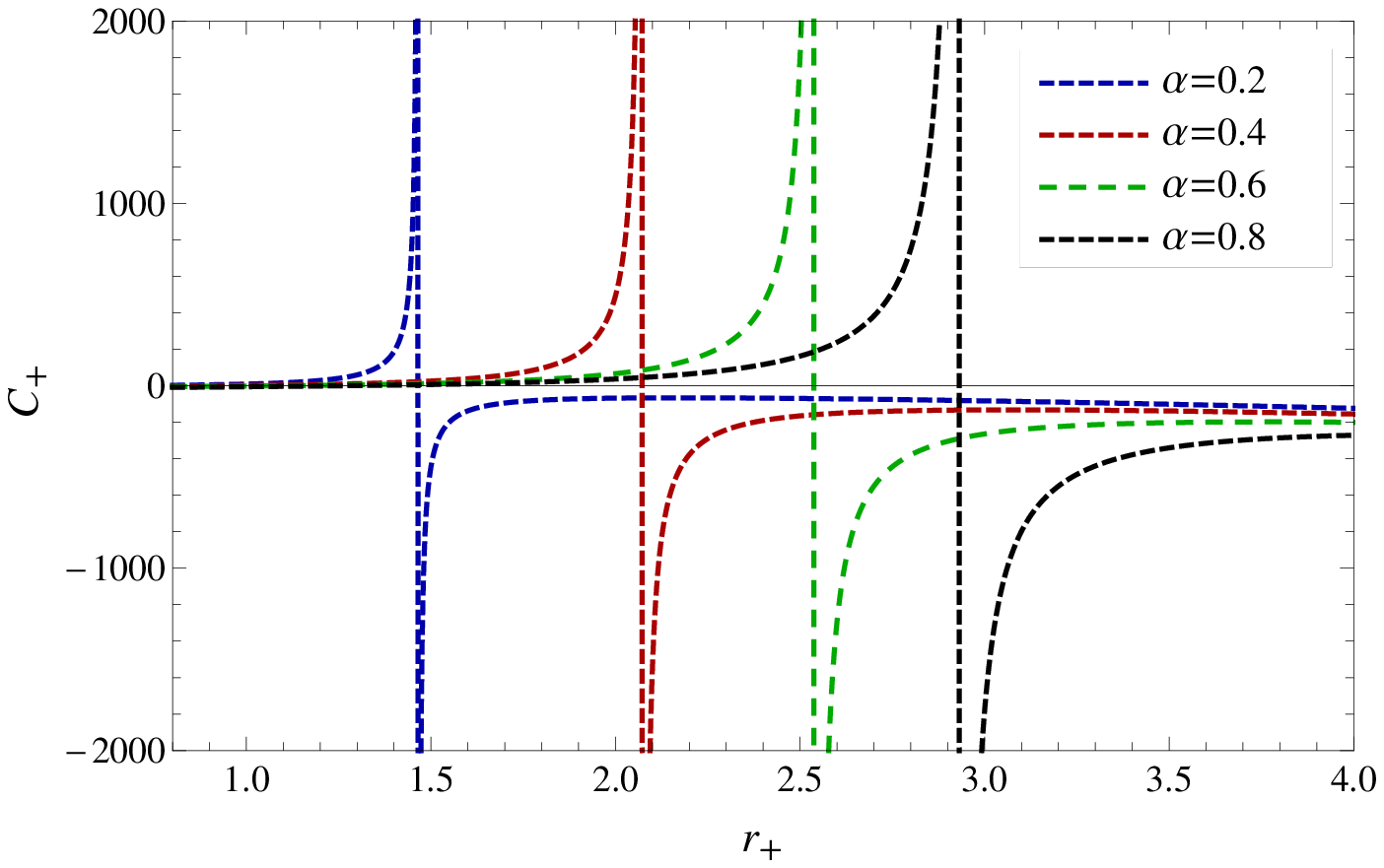}
	\end{tabular}
	\caption{\label{fig:egb:sh} The specific heat  $ ( C_+)$  vs horizon radius ($r_+/\sqrt{\theta}$) for the NC inspired   $4D $ EGB black hole for different values of  $\alpha$ (left), which is compared with the commutative counterpart (right).}
\end{figure*}

Due to the complicated  analytical expression of  the heat capacity, we depict  it in Fig.~\ref{fig:egb:sh} 
for different values of $\alpha$ and compare with the commutative counterpart. 
The heat capacity is discontinuous at $r_+=r_C$ which means that the  second order phase transition happens there \cite{davies}.   It  is positive (negative)
for $r_+ <r_C$ ( $r_+ >r_C$) and thereby suggests the stable (unstable) branch.  The heat capacity diverges  
at critical $r_+=r_C$, where the Hawking temperature attains a maximum value  with $ \left({\partial{T_+}}/{\partial{r_+}}\right) =0$.   
The temperature,  at smaller horizon radius $r_+$ (cf. Fig. \ref{fig:egb:T})   
falls down thereby $\left({\partial{T_+}}/{\partial{r_+}}\right) > 0 $ which implies  
heat capacity $C_{+}>0$ and $C_{+} \to 0$ at smaller horizon radius $r_+$.  At the late stages of the  
black hole  evaporation we have  $C_{+} \ge 0$ (cf. Fig. \ref{fig:egb:sh}).   
Hence the phase transition occurs from a higher mass black hole with negative heat capacity  $C_{+} < 0$ to smaller mass black hole with positive heat capacity $C_{+}>0$. The critical radius $r_C$ depends on the GB parameter $\alpha$ (cf. Fig.~\ref{fig:egb:sh}), such that it increases with $\alpha$. 
The above arguments are valid for both NC inspired black holes and their counterpart (cf. Fig. \ref{fig:egb:sh}).  
However  the critical radii $r_C$  are drastically changed due to NC effects and in this case, 
for a given value of GB parameter $\alpha$, the critical radii $r_C$ are larger than their  
commutative counterparts. It would be interesting  to investigate  how the black hole  with positive specific heat ($C_+>0$) would emerge from thermal radiation through a phase transition. 
Again the large distance limit leads to 
\begin{equation}\label{cegb}
C_{+}^{EGB} = -  \frac{2 \pi r (r_{+}^2 - \alpha) (r_{+}^2 + 2 \alpha)}{2 \alpha^2 + 5 r_{+}^2 \alpha - r_{+}^4},
\end{equation}
which is exactly same as the commutative EGB case \cite{Singh:2020nwo}. 
Clearly the heat capacity  is negative definite for Schwarzschild black holes, which reflects the well-known fact that Schwarzschild black holes get hotter as they radiate energy. But for the noncommutative inspired 4D EGB black holes, this system shows a phase transition, implying the existence of a stable phase where heat capacity is positive.

Finally, regarding the black hole remnant  which  is a plausible candidate for  dark energy \cite{jh}  and also likely to  fathom the information loss puzzle \cite{jp}. The double root  $r_E$ of $f(r)=0$  corresponds to the extremal black hole with degenerate horizon such that  $f(r_E)= f'(r_E)=0$.  We can see clearly that  the two horizons  coincide when $r_-=r_+$  the  temperature vanishes at the degenerate horizon leaving a regular double-horizon remnant with $M=M_C$ given by Eq.~(\ref{MC}).   To be precise, the extremal configuration is depicted in Fig. (\ref{f4d})  mean  horizon degenerate to one  at a minimal non-zero mass $M_C$,  which means when $ M < M_C$ there is no event horizon or no black holes. The existence  mass  $M_C$ can  be elucidate that NC 4D EGB black hole can shrink to as a  de Sitter-like core corresponding a remnant $M_C$.  Further, one can notice from the  Fig. (\ref{f4d}),  the  mass  $M_C$  and radius $r_C$ increases with the parameter $\alpha$.   Thus, we can say that due to Hawking radiation, the temperature reaches a peak at
the final stage of the evaporation and then abruptly drops to zero so with  a stable remnant $M_C$
appeared, e.g., the Hawking  temperature drops zero, for for $\alpha = 0.2\,  \theta,\; 0.4 \, \theta,\;  0.6 \, \theta, \; 0.8 \,  \theta$, respectively  at critical  radius $r_{E} = 3.11708 \; \sqrt{ \theta },\; 3.19874 \; \sqrt{ \theta },\; 3.27121 \; \sqrt{ \theta },\;3.33682 \; \sqrt{ \theta }$, where  horizons degenerate, and we are left with an extremal black hole with remnant $M_C = 1.98493 \; \sqrt{ \theta },\; 2.06147 \; \sqrt{ \theta },\; 2.134671 \; \sqrt{ \theta },\;2.22938 \; \sqrt{ \theta }. $

\section{Discussion}\label{concluding}
There is a widespread belief that the  EGB action can be obtained in the low energy limit of string theory and that the NC structure of the spacetime is one of the exotic outcomes of  string theory, and it provides an effective framework to study short distance spacetime dynamics.  Also, NC structures have been individually explored extensively with an emphasis on the black hole spacetime,  and it is natural for them to make their appearance felt as one approaches high energies.
However, a complete NC theory of gravity is far from existence, so to make progress it is imperative to model  NC effects within the commutative framework.  Some progress in understanding  NC gravity has been made by 
formulating models in which GR is commutative form but the NC geometry leads to a smearing of matter distributions.  
However, there has been not much effort to see NC effects in EGB theory. 

Motivated by this and recent activity in regularising EGB gravity to $4D$, 
we have obtained an exact static spherically symmetric black hole solution in
regularized $4D$ EGB gravity inspired by NC geometry, i.e., the NC counterpart of the 
$4D$ EGB solution (\ref{sol:egb}) which exactly regains $4D$ EGB (\ref{lapse}) and 
$4D$ Schwarzschild solutions in the appropriate limits. It is seen that the new solution
smoothly interpolates between a de Sitter core around the origin and $4D$ EGB (\ref{lapse}) 
at a large distance. The NC inspired $4D$ EGB black hole solution (\ref{sol:egb}), subject to value parameters,
admits horizons which could be two, describing a variety of self-gravitating objects, 
including an extremal black hole with degenerate horizons and a nonextremal black hole with Cauchy and event horizons. 

The thermodynamical quantities associated with NC inspired $4D$ EGB black holes have been analyzed as a function of
both $r_{+}$ and $\alpha$. While much of the thermodynamic properties noticed in NC inspired $4D$ EGB black hole are similar to that obtained in commutative counterpart, some significant NC effects and correction to the previously obtained  $4D$ EGB black holes were discovered. The Hawking temperature, in both cases, 
does not diverge as the event horizon shrinks down; instead, it reaches a maximum value for a critical radius and then drops down to zero. In the NC case this happens for larger values of the radius $r_{+}$ (cf. Fig. \ref{fig:egb:T}) 
and also increases with parameter $\alpha$. The entropy of a black hole, in GR, obeys the area law, but not for $4D$ EGB black holes where it has a logarithmic correction term. 

The heat capacity $C_{+}$, as a function of $r_{+}$ and $\alpha$, can be positive (negative) when the horizon 
radii are small (large) (cf. Fig.~\ref{fig:egb:sh}). The divergence of the specific heat at a critical radius $r_C$, 
a function of the GB parameter $\alpha$, is where the Hawking temperature attains a maximum value. These black holes 
are thermodynamically stable (unstable), in both the theories, with a positive (negative) heat capacity $C_{+}$ 
for the range $ r < r_C$ ($r>r_C$) with an stable (unstable) branch. The heat capacity becomes singular 
at a critical radius of $r_C$ which corresponds to the maximum Hawking temperature, and also 
as expected $C_{+} \to 0$ at the minimum mass as evident from the Figs. \ref{f4d}  and \ref{fig:egb:sh}. 
Further, the specific heat $C_{+}$ of $4D$ EGB black holes behave according to the Hawking-Page phase transition as in the AdS black hole such that $C_{+}$ ranges from infinitely negative to infinitely positive and then down to a finite positive (cf. Fig.~\ref{fig:egb:sh}). The infinite change at $r_C$ 
indicates a thermodynamic behavior of the black hole where the Hawking temperature has a maximum. 
We also observe that a thermodynamically unstable region ($C_+<0$) appears for $r>r_C$. 
The critical radius $r_C$ is also affected by the NC. 

There are many interesting avenues that are amenable for future work from the NC inspired black hole 
(\ref{sol:egb}) of the various regularised $4D$ EGB gravities, which are also valid solutions in gravity with a conformal anomaly. For example,  it will be intriguing to apply these solutions to study effects of 
the higher order curvature in a semi-classical analysis of black hole evaporation. 
The results presented here are the generalization of previous discussions, on the $4D$ black hole, 
in GR and EGB gravity, to a more general setting, and the possibility of a further generalization of 
these results to the more general dynamical case is an interesting problem for future research. 

--

\acknowledgements
S.G.G. would like to thank DST INDO-SA bilateral project DST/INT/South Africa/P-06/2016, S.G.G. also thank SERB-DST 
for the ASEAN project IMRC/AISTDF/CRD/2018/000042 and Rahul Kumar for fruitful discussions and suggestions. S.D.M. 
acknowledges that this work is based upon research supported by the South African Research 
Chair Initiative of the Department of Science and
Technology and the National Research Foundation.

\end{document}